\documentclass[useAMS,usenatbib]{mnras}
\pdfoutput=1

\usepackage{times}

\usepackage{amssymb,amsmath}
\usepackage{tabularx}	
\usepackage{graphicx,color}
\usepackage{hyperref}
\usepackage[usenames,dvipsnames]{xcolor}
\usepackage[OT2,T1]{fontenc}
\usepackage[draft]{todonotes}
\title[Discovery of the GLQ SDSS J1442+4055]{Discovery of the optically bright, 
wide separation double quasar SDSS J1442+4055} 

\author[A. V. Sergeyev et al.]{
A. V. Sergeyev,$^{1,2}$\thanks{E-mail: alexey.v.sergeyev@gmail.com}
A. P. Zheleznyak,$^{1,2}$
V. N. Shalyapin,$^{3,4}$
and L. J. Goicoechea$^{4}$
\\
$^{1}$Institute of Radio Astronomy, Krasnoznamennaya 4, 61002 Kharkov, Ukraine\\
$^{2}$Institute of Astronomy of Kharkov National University, Sumskaya 35, 61022 
Kharkov, Ukraine\\
$^{3}$Institute for Radiophysics and Electronics, National Academy of Sciences 
of Ukraine, 12 Proskura St., 61085 Kharkov, Ukraine\\ 
$^{4}$Departamento de F\'isica Moderna, Universidad de Cantabria, Avda. de Los
Castros s/n, 39005 Santander, Spain
}

\date{Accepted XXX. Received YYY; in original form ZZZ}

\pubyear{2015}

\begin{document}
\label{firstpage}
\pagerange{\pageref{firstpage}--\pageref{lastpage}}
\maketitle

\begin{abstract}
Optically bright, wide separation double (gravitationally lensed) quasars can be 
easily monitored, leading to light curves of great importance in 
determining the Hubble constant and other cosmological parameters, as well as 
the structure of active nuclei and halos of galaxies. Searching for new double 
quasars in the SDSS-III database, we discovered SDSS J1442+4055. This consists 
of two bright images ($r \sim$ 18$-$19) of the same distant quasar at $z$ = 
2.575. The two quasar images are separated by $\sim$ 2\farcs1, show significant 
parallel flux variations and can be monitored from late 2015. We also found 
other two double quasar candidates, SDSS J1617+3827 ($z$ = 2.079) and SDSS 
J1642+3200 ($z$ = 2.264), displaying evidence for the presence of a lensing 
object and parallel flux variations, but requiring further spectroscopic 
observations to be confirmed as lensed quasars.          
\end{abstract}


\section{Introduction}
\label{sec:intro} 

Variability studies of gravitationally lensed quasars (GLQs) are a powerful 
means to study the Universe \citep[e.g.][]{Sch06}. For example, time delays of 
GLQs are used to estimate cosmological parameters, primarily the current 
expansion rate of the Universe \citep[the so-called 
Hubble constant;][]{Ogu07,Suy10,Suy13,Ser14,Rat15}. In addition, 
microlensing-induced flux variations reveal, among other things, the structure 
of quasars and properties of populations of stars (microlenses) in galaxies 
\citep[e.g.][]{Koc04,Eig08,Poi08,Dai10,Mor10,Mor12,Hai13,Mos13,Bla14,Mac15}. 
Multiwavelength intrinsic variability of GLQ images also produces valuable 
reverberation maps of emitting regions of quasars \citep{Gil12,Goi12}, which 
will become in critical tools to check microlensing-derived source structures.

As searching for new lensed quasars and conducting subsequent monitoring 
campaigns are tasks of great relevance in cosmology, current quasar databases 
are being deeply mined to discover new GLQs 
\citep[e.g.][]{Ogu06,Ina12,Jac12,Ely13,Buc14,Agn15a,Nor15}, and several 
techniques are being developed to search for GLQs in ongoing and future surveys 
\citep[e.g.][]{Agn15b,Cha15}. The GLQ zoo includes rare phenomena consisting of 
two optically bright images ($r <$ 20) of the same distant quasar that are 
separated by $\Delta \theta >$ 2\arcsec. Only five out of the 100 
multiply-imaged systems in the CfA-Arizona Space Telescope Lens Survey (CASTLES) 
database\footnote{\url{https://www.cfa.harvard.edu/castles/}} satisfy these 
conditions. Each of the optically bright, wide separation double quasars is of 
particular interest because it can be easily monitored with an 1$-$2m class 
telescope in average seeing conditions, and the time delay between its two 
images can be obtained from light curves and cross-correlation techniques 
\citep[e.g.][]{Poi07,Sha08,Kop12,Eul13,Rat13}. Detection of time delays between 
different optical bands \citep[e.g.][]{Kop10} and uncorrelated 
(microlensing-induced) variations in light curves of the quasar images 
\citep[e.g.][]{Osc13} is also possible. 

The ongoing monitoring projects do not only focus on bright and 
widely-separated double quasars, although several targets belong to this type of 
systems. There are currently many cross-correlation techniques to determine time 
delays of GLQs \citep[e.g.][]{Lia15}, and the Cosmological Monitoring of 
Gravitational Lenses (COSMOGRAIL) is a collaboration targeting $\sim$ 20 GLQs to 
(mainly) measure their time delays \citep{Tew12}. The COSMOGRAIL 
website\footnote{\url{http://cosmograil.epfl.ch}} incorporates a list of main 
publications, light curves and other relevant information. This large 
collaboration uses 1$-$2m telescopes in both hemispheres, and has published 
light curves and time delays of 7 GLQs with optically bright images. Two of them 
are widely-separated doubles \citep{Eul13,Rat13}, three are quads with some 
widely-separated images, and two are doubles with $\Delta \theta \sim$ 1\arcsec. 
Other groups are also involved in similar programmes. For example, the 
Gravitational Lenses and Dark Matter (GLENDAMA) project is conducting an optical 
monitoring of $\sim$ 10 bright GLQs with different angular 
sizes\footnote{\url{http://grupos.unican.es/glendama/LQLM_results.htm}}. The 
GLENDAMA team uses a 2m robotic telescope in the north hemisphere with the 
purpose of measuring time delays, and analysing intrinsic and extrinsic 
(microlensing) flux variations \citep[e.g.][]{Goi10}. 

In this paper, we present our ongoing search for GLQs in the Sloan Digital Sky 
Survey III (SDSS-III) database \citep{Eis11}. The selection of GLQ candidates 
and the corresponding identification campaigns are described in 
Sec.~\ref{sec:selide}. All candidates are composed of two point-like sources: 
one of them is a confirmed quasar and the other belongs to an unknown class. Our
spectroscopic observations confirm the GLQ nature of SDSS J1442+4055, which has 
been independently discovered by \citet{Mor15}. In Sec.~\ref{sec:q1442}, using 
new deep imaging in the $r$ band, we present astro-photometric solutions and 
lens models for this optically bright, wide separation lens system. Our 
conclusions and future prospects appear in Sec.~\ref{sec:final}.

\section{Selection and identification of GLQ candidates}
\label{sec:selide}

\subsection{Samples} 
\label{sec:sample}

Our first sample (S1) of GLQ candidates is based on the SDSS-III Data Release 9 
\citep{Ahn12}. We used the Data Release 9 Quasar (DR9Q) catalog containing 87 
822 spectroscopically confirmed objects \citep{Par12}, as well as $gri$ frames 
including these quasars and other surrounding sources. We initially checked for 
the presence of some point-like source in the vicinity (1$-$6\arcsec\ apart) of 
each DR9Q quasar over the redshift interval 0.8 $< z <$ 4 and brighter than $r$ 
= 21. After finding an association, the similarity of colours of the quasar and 
its companion was probed. Specifically, we estimated the average magnitude 
difference AMD = ($\Delta g$ + $\Delta r$ + $\Delta i$)/3 for each pair, and 
then selected pairs satisfying the constraints d$g <$ 0.4, d$r <$ 0.4 and d$i <$ 
0.4, where d$g$ = $\Delta g$ - AMD, d$r$ = $\Delta r$ - AMD and d$i$ = $\Delta 
i$ - AMD. After visual inspection of the selected targets, we obtained a sample 
of 39 candidates. 

To construct a second sample (S2), we used the SDSS-III DR10  
\citep{Ahn14,Par14} and adopted a selection method different from the previous 
one (see above).  We concentrated on quasars in the range RA = 180$-$360\degr, 
over the redshift interval 1 $< z <$ 5 and brighter than $r$ = 20. We then 
searched for neighbour point-like sources at $\Delta \theta \leq$ 6\arcsec. For 
the widely-separated quasar-companion pairs ($\Delta \theta >$ 2\arcsec), a 
detailed colour test was also done. Using the corresponding $ugriz$ frames, we 
compared colours of quasars and their companions, and selected pairs satisfying 
the constraints $\left|\Delta (u-g)\right| <$ 0.4, $\left|\Delta (g-r)\right| <$ 
0.2, $\left|\Delta (r-i)\right| <$ 0.2 and $\left|\Delta (i-z)\right| <$ 0.4.
After visual inspection, we selected 17 GLQ candidates for the S2 
sample\footnote{To achieve a reasonable balance between compact ($\Delta \theta 
\sim$ 1$-$2\arcsec) and wide separation candidates, we did not consider targets 
with $\Delta \theta >$ 2\arcsec\ in the RA interval 180$-$210\degr}. None of 
these new targets is included in S1.

Our selection criteria differ from those used to make the statistical lens 
sample of the SDSS Quasar Lens Search \citep[SQLS;][]{Ogu06,Ina12}, since we 
focus on double quasar candidates with separations of a few arcseconds, which 
are within a larger cosmic volume and cover a wider range of magnitudes. 
However, the SQLS 
database\footnote{\url{http://www-utap.phys.s.u-tokyo.ac.jp/~sdss/sqls/lens.html}}
also contains additional double quasars that are not included in the statistical 
sample \citep[e.g.][]{Ina09}.  

\subsection{Follow-up observations of selected subsamples} 
\label{sec:subsam} 

\begin{table*}
\centering
\caption{Identification of GLQ candidates. For both subsamples (SS1 and SS2), 
besides the name, redshift and position (J2000) of the quasars, it is also shown 
the separation between each quasar and its previously unidentified companion, as 
well as the follow-up observations. The last column informs about our main 
results using deep imaging and/or spectroscopy.}
\label{tab:selide}
\begin{tabularx}{\textwidth}{cccccccc} 
\hline
Quasar & $z$ & RA (\degr) & DEC. (\degr) & $\Delta \theta$ (\arcsec) & \multicolumn{3}{c}{Follow-up observations}\\
 & & & & & Imaging (Date) & Spectroscopy (Date) & GLQ\\
\hline
\\
\multicolumn{8}{c}{SS1}\\
\\
SDSS J0734+2733 & 1.918 & 113.52813 & +27.56545 & 2.55 & & LT/FRODOSpec (2014 Dec 20) & No\\
SDSS J0735+2036 & 2.189 & 113.89411 & +20.60950 & 3.68 & & TNG/DOLORES (2014 Mar 02) & No\\
SDSS J0755+1400 & 0.845 & 118.83309 & +14.01166 & 2.14 & NOT/ALFOSC (2014 Mar 21) & & Unlikely\\
\\
\multicolumn{8}{c}{SS2}\\
\\
SDSS J1239+4447 & 2.037 & 189.76377 & +44.78371 & 1.77 & AZT-22/SNUCam (2015 Jun 05) & & Unlikely\\
SDSS J1410+2340 &	2.449 & 212.51833 & +23.68192 & 3.58 & AZT-22/SNUCam (2015 Jun 07) & & Unlikely\\
SDSS J1442+4055 &	2.575 & 220.72826 & +40.92655 & 2.07 & AZT-22/SNUCam (2015 Jun 10) & LT/SPRAT (2015 Aug 29) & Yes\\ 
SDSS J1523+1220 &	2.403 & 230.81149	& +12.33632	& 2.19 & AZT-22/SNUCam (2015 Jun 17) & & Unlikely\\
SDSS J1526+0151 &	3.167 & 231.63047 & +01.85088 & 1.41 & AZT-22/SNUCam (2015 Jun 11) & & Unlikely\\
SDSS J1554+2616 &	2.321 & 238.54804 & +26.27660 & 2.16 & AZT-22/SNUCam (2015 Jun 18) & & Unlikely$^{\star}$\\
SDSS J1600+1028 &	1.535 & 240.06076	& +10.48122 & 1.28 & AZT-22/SNUCam (2015 Jun 16) & & Unlikely\\
SDSS J1617+3827 & 2.079 & 244.47052 & +38.46030 & 2.12 & AZT-22/SNUCam (2015 Jun 15) & & Yes?\\
SDSS J1619+1621 &	2.426 & 244.87857 & +16.35632 & 2.63 & AZT-22/SNUCam (2015 Jun 18) & & Unlikely\\
SDSS J1642+3200 & 2.263 & 250.71282 & +32.00811 & 2.91 & AZT-22/SNUCam (2015 Jun 11) & & Yes?\\
SDSS J1648+3400 &	2.144 & 252.24318 & +34.01664 & 1.37 & AZT-22/SNUCam (2015 Jun 07) & & Unlikely\\
SDSS J1655+1948 & 3.261 & 253.93153 & +19.81310 & 3.85 & & LT/SPRAT (2015 September 10) & No\\ 
SDSS J2053-0100 &	2.207 & 313.38756 & $-$01.01622 & 1.43 & AZT-22/SNUCam (2015 Jun 17) & & Unlikely\\
SDSS J2153+2732 &	2.216 & 328.31784 & +27.54302 & 3.62 & AZT-22/SNUCam (2015 Jun 20) & & Unlikely\\
\hline
\\
\end{tabularx}
\begin{flushleft}
$^{\star}$ Identified as a quasar+star system by \citet{Mor15}.
\end{flushleft}
\end{table*}

From the whole S1 sample in Sec.~\ref{sec:sample}, we took a subsample (SS1) of 
three targets to perform follow-up observations (see Table~\ref{tab:selide}). We
focused on $\sim$ 10\% of the initial sample in the RA interval 110$-$120\degr.
These three SS1 candidates in Table~\ref{tab:selide} have a large separation, 
and the inmediate goal of the observations was to check the efficiency of our 
first selection method and try to discover some new GLQ. We attempted to obtain 
decent spectra for all SS1 targets in 2014. However, the spectra of the faintest 
candidate were very noisy, so it was also imaged. For this candidate, in 
Sec.~\ref{sec:result}, we only present the analysis of its astro-photometric 
data. 

We selected $\sim$ 80\% of the S2 sample, i.e., 14 out of the 17 initial 
candidates (see Table~\ref{tab:selide}), to make a second subsample in the RA 
interval 180$-$330\degr\ (SS2). We first obtained follow-up images for 13 SS2 
candidates at the Maidanak Astronomical Observatory (MAO) in 2015 June, and then 
attempted to conduct spectroscopic observations of each target whose follow-up 
images showed evidence for both a lensing galaxy and a variable companion. As 
almost all (12) SS2-MAO targets have a quasar at $z >$ 2, the presence of a 
residual object (residual signal above the noise level) was considered as 
evidence for a lensing galaxy (see Sec.~\ref{sec:result}). In addition, 
companion sources with flux variations (absolute differences between their SDSS 
and MAO magnitudes in the $r$ band) of $\geq$ 0.1 mag were identified as 
variable objects. Unfortunately, we could not complete the spectroscopic 
observations due to observing constraints. However, we obtained spectra for the 
only SS2 target that was not imaged during the intensive observational campaign 
in 2015 June. This last candidate had weaker observing constraints than the 
pending targets showing flux variations and residual sources.

\subsection{Results} 
\label{sec:result}

\begin{figure*}
\centering
\includegraphics[width=\textwidth]{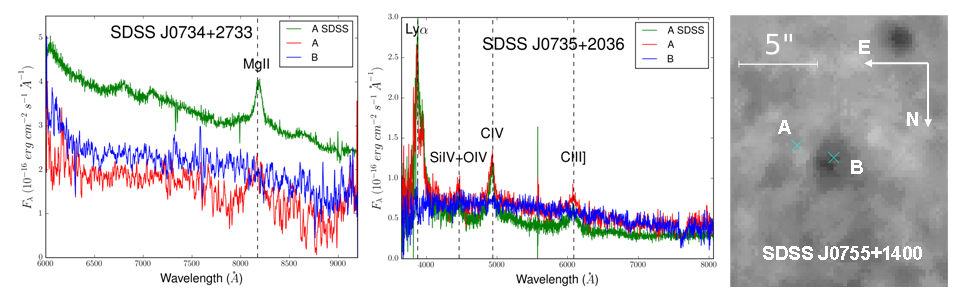}
\caption{Follow-up observations of SS1 candidates. Left and middle: 
spectra of the two components (A and B) of SDSS J0734+2733 and SDSS J0735+2036. 
In each panel, we also show the SDSS spectrum of the quasar (A) for comparison 
purposes. Right: $I$-band residual frame of SDSS J0755+1400. This is an inverted 
greyscale image, so brighter sources correspond to darker zones. Here, B is the 
confirmed quasar and A its companion.}
\label{fig:candi1}
\end{figure*}

SDSS J0734+2733 is the brightest candidate, i.e., two components with $r \sim$ 
17. This was observed with the integral field spectrograph (FRODOSpec) on the 
2m Liverpool Telescope (LT). A 3200 s exposure on a dark night under normal 
seeing conditions was taken with this instrument. FRODOSpec is characterised by 
a field of view and a spatial scale of $\sim$ 10\arcsec$\times$10\arcsec\ and 
0\farcs8 pixel$^{-1}$, respectively, and allows us to extract individual spectra 
for two point-like sources separated by a few pixels \citep{Sha14}. We also note 
that the spectrograph has two independent arms, allowing simultaneous 
spectroscopy at blue and red wavelengths. However, the sensitivity of its red 
arm is significantly higher, so we only show the red spectra of SDSS J0734+2733 
in the left panel of Fig.~\ref{fig:candi1}. The original FRODOSpec/red-arm data
(spectral resolving power $R \sim$ 2200) have been smoothed using a rectangular 
smoothing window of 24 \AA. While the 
confirmed quasar (component A; red line) shows a \ion{Mg}{ii} emission line near 
8200 \AA, this feature is absent in the spectrum of its companion (component B; 
blue line). Hence, we conclude that SDSS J0734+2733 is not a double quasar.

SDSS J0735+2036 was observed with the Device Optimized for the Low Resolution 
(DOLORES) on the 3.6m Telescopio Nazionale Galileo (TNG). We used the LR-B grism
($R \sim$ 600 and 0\farcs25 pixel$^{-1}$) and the 1\farcs5 long slit, putting 
this slit in the direction joining both optically bright components ($r \sim$ 
19). The spectroscopic exposure of 1200 s was done on a dark night. In the 
middle panel of Fig.~\ref{fig:candi1}, we depict the TNG/DOLORES spectra of the 
quasar (component A; red line) and the previously unidentified object (component 
B; blue line). The continuum of B is similar to that of the quasar, but there 
are no emission lines superimposed. Thus, the companion object is likely a star. 

Regarding SDSS J0755+1400 ($r \sim$ 19$-$20), we obtained $I$-band exposures 
with the Andalucia Faint Object Spectrograph and Camera (ALFOSC) on the 2.5m 
Nordic Optical Telescope (NOT). This camera has a spatial scale of 0\farcs19 
pixel$^{-1}$, which makes it a suitable instrument, in good seeing conditions, to 
try to detect a possible lensing galaxy between the two sources. Unfortunately, 
only half of individual frames were useful, and these were combined to produce a 
single deep exposure (1800 s of dark time under good seeing conditions). We then 
subtracted two point-spread functions (PSFs) from the NOT/ALFOSC combined frame 
of the candidate, using nearby stars as PSF templates. In the right panel of 
Fig.~\ref{fig:candi1}, we display the residual frame after subtracting the two 
point-like sources centred at the green crosses. Actually the residual signal 
was smoothed with a median filter in circular neighborhoods (4-pixel radius), 
and contains a possible very faint source ($I >$ 22) close to the quasar B. 
However, this source (if real) is not expected to be the lensing galaxy of a 
quasar at $z$ = 0.845 because so faint lensing objects are usually associated 
with GLQs at $z >$ 3 (see the CASTLES database$^1$). Assuming an early-type lens 
galaxy, we can also infer the expected value of the lens magnitude in the $I$ 
band as a function of the lens redshift $z_{\rm lens}$ \citep[e.g.][]{Rus03}. 
The NOT/ALFOSC data are against the GLQ nature of SDSS J0755+1400, since the 
lens galaxy should have $I <$ 19 at $z_{\rm lens} \leq$ 0.8.

\begin{figure}
\centering
\includegraphics[width=\columnwidth]{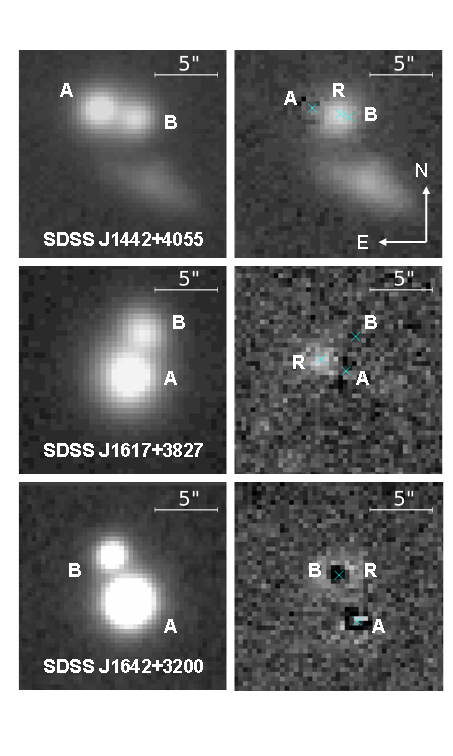}
\caption{MAO $r$-band observations of three SS2 candidates. Left: combined 
3600-s exposures of the quasar-companion (A-B) pairs. Right: residual frames. 
Each frame includes one residual object (R).}
\label{fig:candi2}
\end{figure}

Using the Seoul National University Camera (SNUCam) on the 1.5m AZT-22 Telescope 
(0\farcs27 pixel$^{-1}$) at the MAO, we obtained 3600 s (6$\times$600 s) 
exposures in the SDSS $r$ and/or $i$ bands for 13 out of the 14 SS2 candidates 
(see Table~\ref{tab:selide}). Most individual frames were taken under good 
seeing conditions (FWHM $\leq$ 1\arcsec), and only three combined $r$-band 
frames led to residual signals above the noise level, i.e., residual sources 
with $r \leq$ 23. These combined frames are shown in the left panels of 
Fig.~\ref{fig:candi2}. After subtracting two PSFs within the regions containing 
quasar-companion (A-B) pairs, the residual frames are also presented in the 
right panels of Fig.~\ref{fig:candi2}. We only focused on the three targets 
having an associated residual source (R), which was interpreted as evidence of 
the existence of a lensing galaxy. The average redshift and separation for the 
13 candidates are $\langle z \rangle$ = 2.3 and $\langle \Delta \theta \rangle$ 
= 2\farcs2, so there is a weak constraint $r \leq$ 23 on $r$-band magnitudes of 
possible early-type lens galaxies at $z_{\rm lens} \leq$ 2. We also checked 
the variability of the 
companion sources (comparing their new MAO magnitudes with the corresponding 
SDSS data), since quasars are variable objects. This kind of test has been 
proposed by \citet{Koc06} and \citet{Buc14} among others. The 3 companions are 
variable sources because we detected appreciable changes in their $r$-band 
magnitudes, i.e., $\left|r_{\rm SDSS} - r_{\rm MAO}\right| \geq$ 0.1.  

SDSS J1442+4055 has two bright components ($r \sim$ 18$-$19), and the residual 
object appears in the expected position for a main lensing galaxy. A possible 
secondary lens also lies southwest of this object in the top right panel 
of Fig.~\ref{fig:candi2}. Through PSF photometry in the $r$ band (using standard 
{\small IRAF}\footnote{IRAF is distributed by the National Optical Astronomy 
Observatories, which are operated by the Association of Universities for 
Research in Astronomy, Inc., under cooperative agreement with the National 
Science Foundation} tasks), we calculated magnitude differences betwen two 
epochs separated by $\sim$ 10 years: SDSS(2003 March) $-$ MAO(2015 June). 
For field stars, these differences were only about $\pm$ 0.01, while we obtained 
highly significant flux decrements of $\sim$ 0.1$-$0.2 magnitudes for both 
components. These are strong arguments in favour of the GLQ nature of SDSS 
J1442+4055, and thus, we selected this target for spectroscopic follow-up 
observations with the highest priority. The other two candidates: SDSS 
J1617+3827 ($r \sim$ 19$-$21) and SDSS J1642+3200 ($r \sim$ 18$-$21), do not 
show significant residues on the lines 
joining their components (see the middle and bottom right panels of 
Fig.~\ref{fig:candi2}). The residual objects are located northeast of the bright 
quasar with $r \sim$ 19 (SDSS J1617+3827) or around the faint companion object 
with $r >$ 20 (SDSS J1642+3200). Additionally, the quasars and their companions 
have experienced flux decrements that exceed the noise level (0.01 mag) in 
factors from 3 to 30. Therefore, these two candidates also deserve more 
attention, and they were selected as secondary targets for spectroscopy. 
Unfortunately, we were not able to get decent spectra for both targets.

\begin{figure}
\centering
\includegraphics[width=\columnwidth]{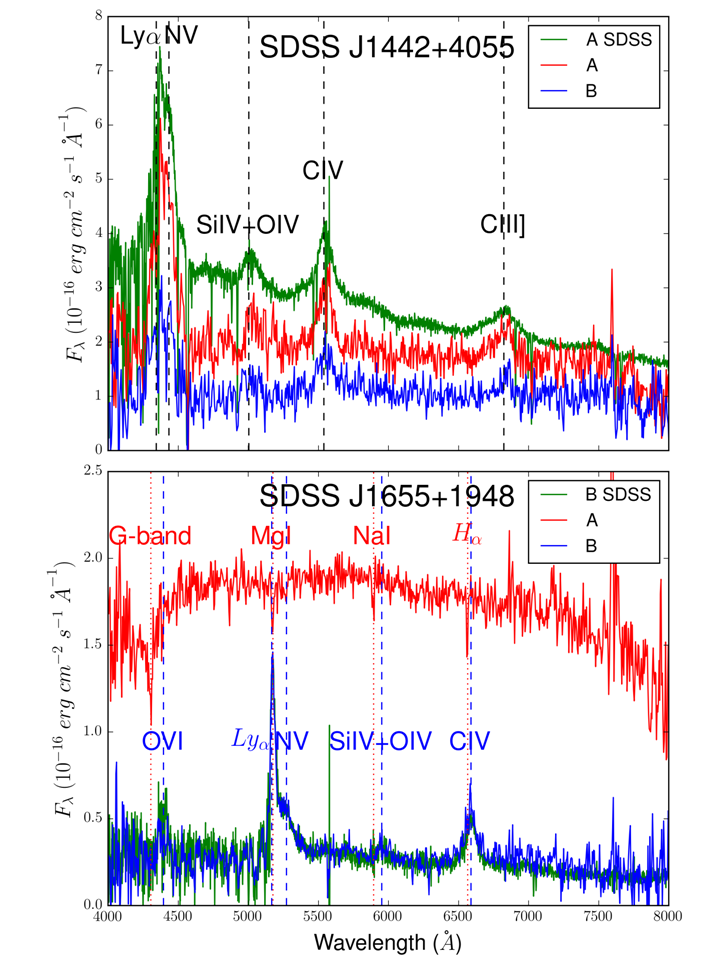}
\caption{LT/SPRAT spectroscopic observations of two SS2 targets. Top: spectra of 
the two components (A and B) of SDSS J1442+4055. The SDSS spectrum of the quasar 
(A) is also displayed in this panel. Bottom: spectra of the two components (A 
and B) of SDSS J1655+1948. We also show the SDSS spectrum of the quasar (B).}
\label{fig:candi2spec}
\end{figure}

We observed SDSS J1442+4055 with the Spectrograph for the Rapid 
Acquisition of Transients (SPRAT) on the 2m LT. Although the sky was relatively 
bright, the strong evidence for a GLQ and the near end of the target visibility 
window justified urgent spectroscopic observations. Using the blue grating and 
the 1\farcs8 long slit ($R \sim$ 350 and 0\farcs44 pixel$^{-1}$), we took five 
600 s exposures in normal seeing conditions (FWHM $\sim$ 1\farcs3). The top 
panel of Fig.~\ref{fig:candi2spec} shows the combined (3000 s) spectra of A 
(quasar) and B (companion source), which incorporate Ly$\alpha$+\ion{N}{v}, 
\ion{Si}{iv}+\ion{O}{iv}], \ion{C}{iv} and \ion{C}{iii}] emission lines at very 
similar wavelengths. A cross-correlation between the two spectra proved that 
their redshifts are identical\footnote{More properly, the redshift difference is 
$\Delta z <$ 0.001, where $\Delta z$ = 0.001 corresponds to a pixel along the 
dispersion axis}, and this corroborated the GLQ nature of SDSS J1442+4055.

We also obtained SPRAT spectra for the only SS2 target that was not observed at 
the MAO. The candidate SDSS J1655+1948 consists of a relatively faint quasar (B) 
with $r \sim$ 20 and a bright companion (A) with $r \sim$ 18, and its 3000 s 
spectra are depicted in the bottom panel of Fig.~\ref{fig:candi2spec}. For this 
GLQ candidate, the A component is a star whose spectrum incorporates several 
absorption lines at $z \sim$ 0. We failed again in a positive identification of 
a candidate selected by standard methods (see Sec.~\ref{sec:sample}), and this 
stresses the efficiency in detecting GLQs using more sophisticated selection 
criteria, i.e., existence of a residual object and variable components.

\section{SDSS J1442+4055} 
\label{sec:q1442} 

\begin{figure}
\includegraphics[width=\columnwidth]{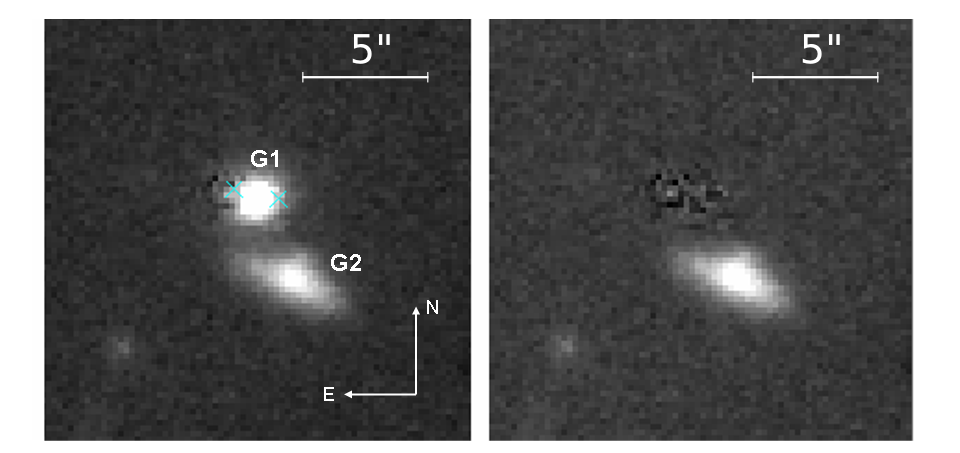}
\caption{{\small GALFIT} model of SDSS J1442+4055 in the $r$ band. This model 
includes three ingredients: two PSFs (quasar images) and the convolution of a 2D 
S\'ersic profile with the PSF (main lensing galaxy G1). We show the residual 
signal after subtracting only the quasar images (left panel) and after 
subtracting the global solution (right panel).}
\label{fig:q1442}
\end{figure}

The MAO $r$-band imaging of SDSS J1442+4055 (FWHM $\sim$ 1\arcsec) was also used 
to infer astro-photometric solutions for the new lens system. Through the 
{\small GALFIT} software \citep{Pen02}, this was modelled as two PSFs (quasar 
images) plus a standard galaxy profile convolved with the PSF. We considered a 
2D S\'ersic profile \citep{Ser63} for the main lensing galaxy, since it can 
describe light distributions for different types of galaxies. In the left and 
right panels of Fig.~\ref{fig:q1442}, we display the residual signal after 
subtracting the two quasar images and the final residues after subtracting the 
global best fit (quasar images + main lensing galaxy), respectively. The main 
lens (G1) and the secondary lens (G2) are clearly resolved in the left panel 
of Fig.~\ref{fig:q1442}. Apart from the solution leaving the concentration index 
($n$) as a free parameter (S\'ersic profile), we also used a de Vaucouleurs 
profile for G1 to obtain a second solution, i.e., fixing $n$ = 4, rather than 
leaving it as a free parameter (see Table~\ref{tab:q1442}). This is a reasonable 
profile for early-type galaxies and galactic bulges, and it may be useful for 
extracting quasar fluxes from optical frames in monitoring campaigns 
\citep[e.g.][]{Sha08}. However, the reduced chi-squared value is 1.8, so the 
1$\sigma$ confidence intervals that we give in Table~\ref{tab:q1442} should be 
taken with caution.

\begin{table*}
\centering
\caption{Astro-photometric solution for SDSS J1442+4055 in the $r$ band. Here, 
($x_{\rm{B}}$, $y_{\rm{B}}$) is the position of B, and the last six parameters 
correspond to G1 (de Vaucouleurs profile): position ($x_{\rm{G1}}$, 
$y_{\rm{G1}}$), magnitude ($r_{\rm{G1}}$), effective radius $r_{\rm{eff}}$, 
ellipticity $e$, and position angle of the major axis $\theta_e$ (it is measured 
east of north).}
\label{tab:q1442}
\begin{tabular}{cccccccc} 
\hline
$x_{\rm{B}}$ (\arcsec) & $y_{\rm{B}}$ (\arcsec) & $x_{\rm{G1}}$ (\arcsec) & 
$y_{\rm{G1}}$ (\arcsec) & $r_{\rm{G1}}$ (mag) & $r_{\rm{eff}}$ (\arcsec) & $e$ & 
$\theta_e$ (\degr)\\
\hline
2.096 $\pm$ 0.003 & $-$0.506 $\pm$ 0.001 & 1.342 $\pm$ 0.013 & $-$0.344 $\pm$ 0.008 
& 19.47 $\pm$ 0.01 & 0.59 $\pm$ 0.02 & 0.19 $\pm$ 0.04 & $-$12 $\pm$ 6\\
\hline
\end{tabular}
\end{table*}

SDSS J1442+4055 has been independently discovered by \citet{Mor15}. When we 
were writing this report, \citet{Mor15} announced the discovery of thirteen 
GLQs, including SDSS J1442+4055 between their confirmed lens systems. They used 
a 900 s $i$-band exposure with the 2.4m Hiltner Telescope to model the system,
whereas our fits were based on an $r$-band combined frame with an equivalent 
exposure time (2.4m telescope) of 1400 s. We note that our estimation of 
$r_{\rm{G1}}$ in Table~\ref{tab:q1442} coincides with the ($g_{\rm{G1}}$ + 
$i_{\rm{G1}}$)/2 value from Table 3 of More et al. Although there is also 
agreement in the values of $e$ and $r_{\rm{eff}}$ (the S\'ersic profile led to 
$r_{\rm{eff}} \sim$ 1\arcsec\ and a large index $n >$ 4), our $\theta_e$ values 
are separated by $\sim$ 90\degr\ from the More at al.'s solution for the 
position angle.

Even though the redshift of the main lensing galaxy (G1) is still unknown, 
the current SDSS-MAO data give a hint as to the $z_{\rm lens}$ value. The SDSS 
spectrum of the A image shows the existence of an absorption system at $z_{\rm 
abs}$ = 1.946, which was identified from 12 different absorption features in the 
wavelength range 4500$-$8500 \AA. However, the Rusin et al.'s scheme 
\citep{Rus03} is in clear disagreement with a bright lens ($r \sim$ 19.5; see 
Table~\ref{tab:q1442}) at $z_{\rm lens} \sim$ 2, so the distant absorption 
system is most likely not linked to G1. We may also consider a second scenario 
where G1 is physically associated with G2 ($r \sim$ 20). As the SDSS database 
includes a 
photometric redshift for G2 (0.32 $\pm$ 0.05), $z_{\rm lens}$ would be $\sim$ 
0.3$-$0.4. This lens redshift is in good agreement with the MAO brightness of 
G1, and thus, the gravitational lens could consist of two galaxies at similar 
moderate values of $z$. 

We also modelled SDSS J1442+4055 using the relative astrometry in 
Table~\ref{tab:q1442} and the $i$-band flux ratio by \citet{Mor15}, as well as a
singular isothermal sphere (SIS) mass model in a standard cosmology 
\citep[e.g.][]{Sch06}. The {\small LENSMODEL} software \citep{Kee01} led to a 
solution with $\chi^2 \gg$ 1, indicating we need to improve our model. The 
simplest option is to add ellipticity to the SIS mass distribution of G1. Thus, 
we used the {\small LENSMODEL} package to find the singular isothermal ellipsoid 
(SIE) solution for the astro-photometric constraints. The number of model 
parameters was the same as the number of observational constraints, and the new 
solution was characterized by a mass scale $b \sim$ 1\farcs08, an ellipticity $e 
\sim$ 0.017, a mass orientation of about $-$60\degr\ and $\chi^2 \sim$ 0. 
Although the ellipticity and orientation of the mass do not coincide with the 
ellipticity and orientation for the light distribution of G1 (see the two last 
columns in Table~\ref{tab:q1442}), we remark that we are dealing with an 
effective mass distribution, since the secondary lens (G2) was not modelled in 
any way. For this effective lens model (SIE), the expected time delay is $\sim$ 
40 days if $z_{\rm lens}$ = 0.4 (A is leading).  

\section{Conclusions and future work}
\label{sec:final} 

Using the SDSS-III DR9 \citep{Ahn12,Par12} and DR10 \citep{Ahn14,Par14}, we have 
searched for new double quasars with separations of 1$-$6\arcsec. As expected, 
samples of a few candidates selected from standard criteria (i.e., a quasar and 
a point-like companion source, having both similar colours) were largely 
inefficient tools to find new GLQs \citep[e.g.][]{Ina12}. For this reason, we 
introduced two additional ingredients in the selection procedure: presence of a 
residual source (lensing galaxy?) in a new deep exposure, and detection of flux 
variations by comparing the SDSS imaging and the new non-SDSS exposure.

After an observational campaign at the Maidanak Astronomical Observatory, we 
found three GLQ candidates satisfying all our selection criteria. The most 
promising target (SDSS J1442+4055) was spectroscopically observed in 2015 
August, showing two identical spectra for the quasar and its companion 
\citep[see also][]{Mor15}. This is an optically bright, wide separation double 
quasar whose images appreciably vary in parallel on a 10-year timescale (see 
Sec.~\ref{sec:result}). Therefore, SDSS J1442+4055 is an excellent target to be 
monitored with an 1$-$2m class telescope from late 2015. Besides light curves of 
the two quasar images, new spectroscopy and deep imaging are required to 
accurately determine the lens redshift $z_{\rm lens}$ and the astro-photometric 
parameters. The current astro-photometric data suggest that $z_{\rm lens} 
\sim$ 0.3$-$0.4, where the main lens would be the galaxy G1, while another 
neighbor galaxy (G2) would also contribute to the gravitational mirage.
Spectroscopic observations of the other two GLQ candidates (SDSS J1617+3827 and 
SDSS J1642+3200) are also needed to determine their GLQ nature.

The new lens system SDSS J1442+4055 is now included in the GLENDAMA list of 
GLQs, so it will be observed with the LT, the 10.4m Gran Telescopio Canarias and
other facilities. We also plan to perform a spectroscopic follow-up of the two
variable targets with residual sources (SDSS J1617+3827 and SDSS J1642+3200), as 
well as two additional variable targets that do not show residual sources with 
$r \leq$ 23 (SDSS J2053-0100 and SDSS J2153+2732; see Table~\ref{tab:selide}). 
In the near future, we will also explore the feasibility to find new GLQs in the
SDSS-III and other databases, by using pairs of unidentified point-like sources 
showing evidence for variability \citep[e.g.][]{Koc06,Buc14}. 

\section*{Acknowledgements}

We wish to thank the anonymous referee and Anupreeta More for useful suggestions 
and comments that have helped to improve the original manuscript. 
The observations at the Maidanak Astronomical Observatory were supported by the 
ISON project. We thank Seoul National University for equipping the AZT-22 
Telescope with a modern CCD camera. This report is also based on observations 
made with the Liverpool Telescope (Prog. XCL04BL2), the Nordic Optical Telescope 
(Prog. SST2014-057) and the Telescopio Nazionale Galileo (Prog. A28DDT8), 
operated on the island of La Palma by the Liverpool John Moores University (with 
financial support from the UK Science and Technology Facilities Council), the 
Nordic Optical Telescope Scientific Association and the Fundación Galileo 
Galilei of the INAF (Istituto Nazionale di Astrofisica), respectively, in the 
Spanish Observatorio del Roque de los Muchachos of the Instituto de 
Astrof\'{\i}sica de Canarias. We thank the staff of all used telescopes for a 
kind interaction before, during and after the observations. We also used data 
taken from the SDSS-III web site (\url{http://www.sdss3.org/}), and we are 
grateful to the SDSS-III collaboration for doing that public database. This 
research has been supported by the Spanish Department of Research, Development 
and Innovation grant AYA2013-47744-C3-2-P (GLENDAMA project), and the University 
of Cantabria.







\bsp	
\label{lastpage}
\end{document}